# Recovery Swaps

**Arthur M. Berd**
Lehman Brothers Inc.

*We derive an arbitrage free relationship between recovery swap rates, digital default swap spreads and conventional CDS spreads, and argue that the fair forward recovery rate used in recovery swaps must contain a convexity premium over the expected recovery value.*

## RECOVERY SWAP CONTRACTS

Recovery swaps are the most recent innovation in the credit derivatives market. In this contract, two counterparties agree to exchange the realized recovery vs. the preset recovery value (recovery swap rate) in case of default, with no other payments being made in any other scenario. The typical recovery swaps have no running or upfront payments of any kind. In keeping with the swaps nomenclature, the counterparties are denoted as the "payer" and "receiver" of the realized recovery rate, correspondingly. From a payer's perspective, the payoff will be positive when the realized recovery is less than the preset swap rate level.

$$[1] \quad \text{Payoff} = \begin{cases} R_{\text{swap rate}} - R_{\text{realized}} & [\text{in case of default prior to maturity}] \\ 0 & [\text{otherwise}] \end{cases}$$

Recovery swaps allow investors to eliminate the uncertainty of the future recovery payment which is present in the conventional (i.e. floating recovery) CDS, whose protection payment in case of default depends on the post-default price of the reference obligation. In a sense, one might also call these contracts "recovery forwards" because they operate similarly to, for example, an FX forward contract which allows investors to eliminate the uncertainty about the future FX rate. Such a name would also emphasize the fact that their payoff is linear in terms of the future realized (i.e. "spot") recovery rate. However, we will follow the lingo which is closer to the CDS market and continue calling these contracts "recovery swaps".

The market for recovery swaps is driven primarily by the hedging needs of synthetic CDO investors and other structured credit market participants. The reason why they need recovery swaps is that the pricing and risk management of these more complex securities depends separately on the default event risk and recovery risk. This is in contrast with the on-the-run (par) conventional CDS contracts whose value depends on the combination of the two risk factors – the default loss risk.

Consider, for example, a synthetic CDO tranche. Most models that price these tranches must rely on some assessment of the joint risk-neutral default probability between various underlying names. This joint default probability is a non-linear function of individual default probabilities of each name and some measure of their correlation, usually incorporated via a copula-based framework.

The individual implied default probabilities must be calibrated to the observed conventional CDS spreads. It is this calibration which induces a strong dependence between the assumed recovery rate and the single-name implied default probability. The relationship between the two can be loosely summarized by the so-called "credit triangle" formula:

$$[2] \quad \text{CDS Spread} \approx \text{Default Loss Risk} = \text{Default Probability} \times (1 - \text{Recovery})$$

When calibrated to a given observed spread level, the implied default probability will be higher for higher values of assumed recovery rate. This implied default probability will then become an input into complex correlated default calculations used to price the synthetic CDO tranche. The same recovery value will be applied in the very end when figuring out the loss distribution on the tranche, but the dependence has been broken – the loss distribution on the tranche will no longer depend just on the default loss ratio, but separately on each individual default probability and each recovery rate. Consequently, the tranche values and risk measures also depend both explicitly, and implicitly, via calibration, on the recovery rate assumptions for underlying names.

In fact, this dependence is not unique to the tranche products, but is generally the case for most structured credit products and even off-the-run conventional CDS with non-zero mark-to-market and non-par cash credit bonds. In all these cases there are so called "digital" default risks which are characterized by a given fixed dollar loss in case of default. For example, in case of an off-market CDS with positive mark-to-market the current present value on the CDS represents a digital default risk because this unrealized gain will disappear without compensation if the underlying name defaults immediately.

For many years, there existed a small niche in the credit derivatives market which catered to clients who needed to hedge such digital risks. The so-called digital default swaps (DDS) differ from the more conventional liquid CDS by pre-setting contractually the value of the protection payment in case of default. This contractual protection payment is usually quoted in terms of "contractual recovery" rate, which could be set at any number. If it was set at zero, then a protection buyer would get a payment equal to the full notional in case of default. If it was set at 50%, then the protection buyer would get a payment equal to half of the notional.

One of the main uses of DDS was the risk management of corporate accounts receivables, operational risks and other such exposures whose value is not naturally tied to the price of the underlying issuer's bonds or traded loans. Unlike asset managers or banks who need to hedge the losses on a corporate bond or loan portfolios and thus find the conventional floating rate CDS the most natural proxy, the corporate treasurers who need to hedge the specific dollar exposure to a trading counterparty have found the DDS to be better suited (even though often not liquid enough due to a niche nature of that market).

With the advent of the recovery swaps the toolkit for separate hedging of credit risks is becoming more complete:

- **Default loss risk (i.e. full credit risk with mark-to-market exposure):** hedged by conventional (floating recovery) CDS
- **Default event risk:** hedged by digital default swaps, i.e. DDS
- **Recovery rate risk:** hedged by recovery swaps
- **Credit risk without mark-to-market exposure:** hedged by constant maturity default swaps (CMDS)
- **Spread volatility risk:** hedged by (short-term) credit default swaptions

The conventional CDS still remain the linchpin of the credit derivatives market and continue to account for the majority of all traded notionals, according to the most recent survey by the British Bankers Association. But the existence of the expanded toolkit also implies certain connections and (partial) substitution ability between various credit derivatives instruments.

## ARBITRAGE-FREE PRICING OF RECOVERY SWAPS

Recovery swaps can be fully replicated by a combination of conventional and digital CDS. This replication, as usual, implies an arbitrage-free relationship between these three instruments which we will derive below.

Consider a zero-investment composite trade depicted in Figure 1: a payer recovery swap with the fixed swap rate R(swap), and a long-short position where an investor bought a digital default protection with contractual recovery $R_{DDS}$ and sold protection via conventional CDS.

The hedge ratios for the DDS and CDS should be chosen so that the net cash flows are zero in all scenarios. Consider the case of default:

$$[3] \quad CF_{default} = \left(R_{swap} - H_{CDS} + H_{DDS} \cdot (1 - R_{DDS})\right) + (H_{CDS} - 1) \cdot R_{realized}$$

In order to guarantee that these cash flows are equal to zero regardless of the realized recovery rate, one must have:

$$[4] \quad \begin{cases} H_{CDS} &= 1 \\ H_{DDS} &= \dfrac{1 - R_{swap}}{1 - R_{DDS}} \end{cases}$$

Because the cash flows in case when the trades mature without default are identically zero, and because the replication requirement for the case of default uniquely fixes both hedge ratios, then the comparison of the premium cash flows results in an arbitrage-free relationship between the recovery swaps, CDS and DDS.

$$[5] \quad S_{DDS}(R_{DDS}) = \frac{1 - R_{DDS}}{1 - R_{swap}} \cdot S_{CDS}$$

Indeed, if the quoted DDS spread were less (greater) than this value, then this trade would have positive (negative) premium cash flows and zero cash flows otherwise, i.e. there will be an arbitrage.

Recalling that the spread for a DDS with zero contractual recovery encodes the term structure of implied hazard rates and interest, we can interpret equation [5] as a fully consistent generalization of the widely quoted "credit triangle" formula $h = S_{CDS}/(1 - R_{swap})$ for arbitrary term structures of interest and hazard rates.

**Figure 1.** Recovery Swap replication using Digital and conventional CDS

|  | Notional | Premium | Payoff in case of default | Payoff when no default |
|---|---|---|---|---|
| **Payer Recovery Swap** | 1 | - | R(swap)-R | - |
| **Buy: Digital CDS** | H(DDS) | -S(DDS) | 1 - R(DDS) | - |
| **Sell: Conventional CDS** | H(CDS) | +S(CDS) | - (1-R) | - |
| **Net Payments** |  | H(CDS)*S(CDS)-H(DDS)*S(DDS) | R(swap)-R -H(CDS)*(1 – R) + H(DDS)(1 - R(DDS)) | - |

Based on this understanding, it would be natural to think that the recovery swap rates should be used for calibration of the implied hazard rates in the risk-neutral framework. To prove this, assume that we have observed the term structure of CDS spreads and recovery swap rates for a range of maturities [0, T]. According to [5] we can obtain without any further assumptions the term structure of DDS spreads with zero contractual recovery. The pricing of such DDS depends solely on the term structure of risk-neutral survival probabilities (and possibly on the correlation of the default process and the risk-free discount factor).

In a very general form, the pricing of a zero contractual recovery DDS is given by:

$$[6] \quad \frac{1}{1-R_{swap}(t,T)} \cdot S_{CDS}(t,T) \cdot \int_t^T du \cdot E_t\{Z_u \cdot I_{\{u<\tau\}}\} = \int_t^T E_t\{Z_u \cdot I_{\{u<\tau \leq u+du\}}\}$$

Here, we have substituted the no-arbitrage DDS spread in terms of the market-observed CDS spreads and recovery swap from eq. [5]. The variable $\tau$ denotes the (random) default time, $I_{\{X\}}$ denotes an indicator function for a random event $X$, $Z_u$ is the (random) credit risk-free discount factor, and $E_t\{\bullet\}$ denotes the expectation under the risk-neutral measure at time $t$. For convenience in notations, we have adopted an approximation of continuous DDS premium payments (this approximation is irrelevant for the subsequent discussion).

Under the assumption of independence between the default process and the risk-free rates, the expectations of the discount factor and the default indicator factor out. We have $Z(t,u) = E_t\{Z_u\}$, where $Z(t,u)$ is the (non-random) riskless discount factor (i.e. the prices of the zero-coupon credit risk-free bonds). Introducing the term structure of survival probabilities and hazard rates according to the standard reduced-form valuation methodology $Q(t,u) = E_t\{I_{\{u<\tau\}}\} = \exp\left(-\int_t^u h(s) \cdot ds\right)$, we get:

$$[7] \quad \frac{1}{1-R_{swap}(t,T)} \cdot S_{CDS}(t,T) \cdot \int_0^T du \cdot Z(t,u) \cdot Q(t,u) = \int_0^T du \cdot Z(t,u) \cdot h(u) \cdot Q(t,u)$$

One can easily calibrate the term structure of implied hazard rates to the market-observed CDS spreads and recovery swap rates from [7] without making any additional assumptions about the recovery and default process. For example, in the case of flat CDS spreads, recovery swap rates and hazard rates one immediately obtains the "credit triangle" $h = S_{CDS}/(1-R_{swap})$.

We see that the calibration procedure for the implied hazard rates is completely independent of our assumptions regarding the correlation between the recovery rates and default events. While there is plenty of empirical evidence of a negative correlation between default rates and recovery rates (see Altman et. al. [2002]), one can prove from [7] that the calibrated implied hazard rate has the exact opposite behavior – it is a growing function of recovery swap rates, given a level of observed CDS spreads. This is particularly obvious in the case when the credit triangle formula applies.

Note also that in our derivation of the equations [6] and [7] we did not make an assumption about the existence of a liquid DDS market – it is sufficient to assume that the CDS and recovery swaps markets exist, and price the hypothetical DDS based on the known arbitrage-free relationship between the DDS and CDS spreads [5].

As we explained, the recovery swaps are typically traded in a "par" format, i.e. with no upfront or premium payments. The present value of such a swap is zero at inception. However, its present value will generally be not zero after the inception as the market recovery rates fluctuate and move away from the contract fixed rate. Consider a (receiver) recovery swap with a contractual swap rate $R_{swap}$, and assume that the current market rate for the recovery swap of the same maturity is $R_{mkt}$. The present value of this swap is:

$$[8] \quad \begin{aligned} PV_{receiver} &= \int_t^T E_t\{(R_u - R_{swap}) \cdot Z_u \cdot I_{\{u < \tau \leq u+du\}}\} \\ &= \int_t^T E_t\{R_u \cdot Z_u \cdot I_{\{u < \tau \leq u+du\}}\} - R_{swap} \cdot \int_t^T E_t\{Z_u \cdot I_{\{u < \tau \leq u+du\}}\} \\ &= (R_{mkt} - R_{swap}) \cdot \int_t^T E_t\{Z_u \cdot I_{\{u < \tau \leq u+du\}}\} \end{aligned}$$

where $R_u$ is the (random) realized recovery rate in case of default. The last line in this equation is obtained by noting that the first term in the second line is related to the current market recovery rate, which is the rate that would have made the present value equal to zero.

Comparing the last line in [8] with the equation [6] and recalling the definition of the risky PV01 for conventional CDS (see O'Kane and Turnbull [2003]), we finally obtain:

$$[9] \quad PV_{receiver} = \frac{R_{mkt} - R_{swap}}{1 - R_{mkt}} \cdot S_{CDS} \cdot RPV01(S_{CDS}, R_{mkt})$$

Here we have explicitly acknowledged that the RiskyPV01 is a function of the observed current CDS spreads and recovery swap rates since it depends on the implied hazard rates which in turn must be calibrated to $S_{CDS}$ and $R_{mkt}$ as discussed previously. The present value of a payer recovery swap is obviously given by the same formula with an opposite sign. Note that we have not made any assumptions about the dependence between the (risk-neutral) interest rates, realized recovery rates and defaults in deriving the equation [9].

## THE CONVEXITY PREMIUM

Practitioners often use the empirical expected recovery rate (e.g. the long-term average rate of 40%) in calibration of the implied survival probabilities. We now know that it is the market-observed recovery swap rates that must play the role of the risk-neutral recovery parameter when relating the DDS and CDS spreads [5] and when calibrating the implied hazard rates to CDS spreads [7]. However, in the absence of a liquid two-way recovery swap market the price discovery for recovery rates is still in its infancy and one might ask a question about how different should the "fair" recovery swap rate be from the expected average (whether this means expected under real or risk-neutral probability measure).

Equation [9] reveals an important feature of the recovery swaps that might shed some light on the answer to this question. We see that the dependence on the market recovery rate is not linear – the market rate enters not only the difference term in the numerator which reflects the contract payoff structure, but also the denominator and the risky PV01 factor. The latter two dependencies reflect the calibration feedback, which can also be interpreted as a convexity effect if the recovery swap rates are assumed to be equal to expected recovery rates.

Consider what would happen to the mark-to-market of a receiver recovery swap if the market recovery rate dropped while the market CDS rates remained unchanged. The investor would record a negative mark-to-market as she would expect that the (floating) recovery payment will be now less than previously fixed contract recovery rate which she would pay out in case of default. At the same time, the implied default rate calibrated to the unchanged CDS spreads will now be lower, and the expected likelihood of default prior to maturity will also decrease, partially offsetting the loss. Similarly, an upward revision to the expected recovery rate would produce a higher implied default probability and a greater likelihood of a positive payoff.[1] Thus, the investor in a receiver recovery swap benefits from both up and downward changes in the market recovery rate compared to the contractual linear payoff estimate – which is a manifestation of the recovery convexity. Being long such convexity has a positive value, therefore the fair value of the swap rate at which the investor can agree to deal is somewhat higher than the expected future rate. In the rest of the paper we will try to estimate this convexity correction in a heuristic manner.

Let us consider a receiver recovery swap with inception at time $t$, maturity at time $T$ at some intermediate time $t<u<T$ prior to expiry. The (zero) present value of the par recovery swap is equal to the present value of the payoff scenarios when the default happens prior to the time $u$ plus the discounted present value of the remaining "live" recovery swap mark-to-market values averaged over the random market recovery rates $R(u,T)$ and CDS spreads $S_{CDS}(u,T)$ at time $u$. The first portion is simply equal to the present value of a recovery swap with a shorter maturity $u$ estimated at the initial date $t$, which leads us to:

[10]
$$0 = \frac{R_{swap}(t,u) - R_{swap}(t,T)}{1 - R_{swap}(t,u)} \cdot S_{CDS}(t,u) \cdot RPV01(t,u|S_{CDS}(t,u), R(t,u))$$
$$+ \; E_t\left\{ Z_u \cdot \frac{R(u,T) - R_{swap}(t,T)}{1 - R(u,T)} \cdot S_{CDS}(u,T) \cdot RPV01(u,T|S_{CDS}(u,T), R(u,T)) \big| u < \tau \right\}$$

It is difficult to draw a specific conclusion from this relationship in full generality. However, if we make certain simplifying assumptions we may be able to estimate the convexity effect. Let us assume that the interest rates are conditionally independent of the market recovery swap rates and CDS spreads given the survival up to time $u$. Let us also ignore the dependence of the risky PV01 on the level of spreads and recovery rates – both of these dependencies are of second order of magnitude compared to the primary dependence captured by the pre-factors in front of the risky PV01. Finally, let us assume that the term structure of recovery swap rates is flat. Under these assumptions we get the following solution:

[11]
$$R_{swap}(t,T) = \frac{E_t\left\{ \frac{R(u,T)}{1-R(u,T)} \cdot S_{CDS}(u,T) \big| u<\tau \right\}}{E_t\left\{ \frac{1}{1-R(u,T)} \cdot S_{CDS}(u,T) \big| u<\tau \right\}} = \frac{E_t\left\{ R(u,T) \cdot S^0_{DDS}(u,T) \big| u<\tau \right\}}{E_t\left\{ S^0_{DDS}(u,T) \big| u<\tau \right\}}$$

---

[1] As noted in the previous section, these statements are not at all dependent on the assumptions about the correlation between the recovery rate and default events.

The second equation is expressed in terms of zero recovery DDS spreads, using [5].

The equation [11] potentially depends on the time $u$ on the right hand side, but not on the left hand side – this is an artifact of our simplifying assumptions, ignoring small second order effects. What this means is that we must estimate the equation [11] at a "typical" intermediate time before maturity and assume (or verify) that the time dependence is very mild indeed.

One can immediately see from [11] that the relationship between the recovery swap rates and risk-neutral expected recovery rates does in fact strongly depend on our assumptions about the co-movement between the market recovery rates and spreads under the risk-neutral measure. For example, one could make an assumption that the zero-recovery DDS spreads and recovery rates are conditionally independent given survival until $u$, and immediately obtain a simple answer which would of course be fully consistent with the assumption of both flat recovery rate term structure and of mild (or no) dependence on intermediate time $u$.

$$[12] \quad R_{swap}(t,T) = E_t\{R(u,T)|u<\tau\}$$

However, such an assumption would go against the grain of the market behavior as we know it. In fact, the DDS market is practically non-existent and if anything, the DDS spreads are set (not independently discovered) by market makers using the relationship [5] with the liquid CDS spreads and the (estimates of or market values of) recovery rates as primary inputs.

Given this state of affairs, and also the fact the of the three types of securities, the conventional CDS are by far the most liquid and the other two (DDS and recovery swaps) are niche markets with limited price discovery, we believe that it is more meaningful to assume that the CDS spreads are independent of the market recovery swap rates, conditionally on survival until $u$. Under such an assumption, the relationship between the "fair" recovery swap rates today and their expected values in the future is more complex:

$$[13] \quad R_{swap}(t,T) = \frac{E_t\left\{\frac{R(u,T)}{1-R(u,T)}\bigg|u<\tau\right\}}{E_t\left\{\frac{1}{1-R(u,T)}\bigg|u<\tau\right\}} = 1 - \left(E_t\left\{\frac{1}{1-R(u,T)}\bigg|u<\tau\right\}\right)^{-1}$$

If we further assume a stationary distribution of recovery swap rates (which could follow, for example, from a mean-reverting dynamics and would be consistent with the requirement of the mild time dependence), then we can calculate the "fair" recovery swap rate based on the parameters of such a distribution. For example, assuming that the (risk-neutral) distribution of future market recovery rates is close to a normal with a mean $\overline{R}$ and a (relatively narrow) standard deviation $\sigma_R$, we get:

$$[14] \quad R_{swap}(t,T) \approx 1 - \frac{1-\overline{R}}{1+\frac{\sigma_R^2}{(1-\overline{R})^2}} \approx \overline{R} + \frac{\sigma_R^2}{1-\overline{R}}$$

The second term on the right hand side is the amount by which the "fair" recovery swap rate exceeds the expected recovery rate – i.e. the convexity premium.

The convexity premium for the recovery swap rate is intimately related with the DDS digital premium discussed in Berd and Kapoor (2002) where we have argued that the fair DDS spread must be somewhat higher than what one would obtain using the expected (no-premium) recovery rates in [5]. Substituting [14] into [5] we obtain the "fair" DDS spread, which coincides with our estimate in the cited paper[2]:

$$[15] \quad S_{DDS}(R_{DDS}) = S_{CDS} \cdot \frac{1-R_{DDS}}{1-\overline{R}} \cdot \left(1 + \frac{\sigma_R^2}{(1-\overline{R})^2}\right)$$

The uncertainty about the recovery value can be quite high, in the 10-20% range, even when conditioned on the current state of the economy (and much higher unconditionally over long periods of time). For the typical levels of expected recovery rate of senior unsecured debt (around 40%) and recovery uncertainty (around 15%), the estimate for the recovery swap rate is $0.4+0.15^2/(1-0.4)=43\%$, i.e. quite a bit higher than the 40% from which we started. Thus, the convexity premium effect can have a significant impact on the pricing of both recovery swaps and digital default swaps.

Market practitioners recognize the no-arbitrage relationship [5] between the recovery swaps, CDS and DDS, as can be seen from the Figure 2. There, we show recently quoted mid-levels for recovery swap rates, CDS and DDS (with contractual R=0) spreads. We also show the "implied" recovery rate calculated from the comparison of the CDS and DDS spreads using the equation [5]. We can see that the implied recovery swap rate is in line with the quoted recovery swap rate (and is well inside the quoted bid-offer spread).

It is not possible to judge from the Figure 2 whether or not the market recognizes the convexity premium in recovery swap rates, because the "expected recovery rate" is not an observable quantity. However, investors who use such quotes in conjunction with their own estimates of the expected recovery rate (and who agree with our assumptions concerning the dynamics of market recovery rates and CDS spreads) should take the convexity premium effect into account when deciding whether the recovery swap offers relative value. In particular, a recovery swap rate that is higher than the investor's projection of the expected rate may or may not be a sufficient indication of a relative value depending on the estimates of the future uncertainty of the recovery rates, in accordance with [14]. However, a recovery swap rate that is lower than the investor's projection would, in our opinion, constitute such a relative value because we know that the convexity premium is positive. To monetize this opinion, the investor would buy a receiver recovery swap (i.e. pay a lower fixed rate).

**Figure 2. Examples of quoted recovery swap, CDS and DDS rates**

| Ticker | Recovery Swap Rate (%) | CDS (bp) | DDS(R=0) (bp) | Implied R (%) |
|---|---|---|---|---|
| CA | 34.5 | 80 | 122 | 34.8 |
| FMC | 39.5 | 174 | 289 | 39.7 |
| GECC | 38.0 | 26 | 42 | 38.6 |
| T | 37.0 | 242 | 386 | 37.4 |

---

[2] *The relative DDS spread convexity used in Berd and Kapoor (2002) can be estimated from the "no-premium" DDS spread by taking its second derivative with respect to mean recovery rate and leads precisely to [15]*

$$\gamma(\overline{R}) \approx \left(S_{CDS} \cdot \frac{1-R_{DDS}}{1-\overline{R}}\right)^{-1} \cdot \frac{\partial^2}{\partial \overline{R}^2}\left(S_{CDS} \cdot \frac{1-R_{DDS}}{1-\overline{R}}\right)\bigg|_{S_{CDS}=\text{const}} = \frac{2}{(1-\overline{R})^2}$$

In conclusion, we note that if or when recovery swaps become much more liquid, we would regard the market-observed recovery swap rates to be the "true" implied recovery value subsuming the convexity premium effects, and use it for calibration of implied default rates from market-observed CDS spreads – thus bringing the pricing framework for credit derivatives firmly into an arbitrage-free setting. There will be no need to worry about the differences between the fair recovery swap rate and the expected recovery rate.

In fact, when independent price discovery in recovery swap rates and CDS spreads becomes reality, we can test the independence between these rates instead of simply assuming it, and perhaps use the more general version of the equation [11] to estimate the relationship between the fair recovery swap rates and the parameters of the joint distribution of future market recovery rates and CDS spreads. The equation [5] would continue to govern the relationship between the CDS, DDS and recovery swaps.

*Acknowledgments:* I would like to thank Roy Mashal, Lutz Schloegl and especially Marco Naldi for very valuable comments and discussions.

## REFERENCES

**Altman, E. I., B. Brooks, A. Resti and A. Sironi (2002):** *The Link between Default and Recovery Rates: Theory, Empirical Evidence and Implications*, working paper, NYU Stern School of Business

**Berd, A. and V. Kapoor (2002):** *Digital Premium*, Journal of Derivatives, vol. 10 (3), p. 66

**O'Kane, D., and S. M. Turnbull (2003):** *Valuation of Credit Default Swaps*, Quantitative Credit Research Quarterly, vol. 2003-Q1/Q2, Lehman Brothers